\documentstyle[prd,aps]{revtex} 
\draft

\begin{document}
\twocolumn[\hsize\textwidth\columnwidth\hsize\csname
@twocolumnfalse\endcsname
\title{%
\hbox to\hsize{\normalsize To be published in Physical Review D\hfil 
Preprint MPI-PTh/96-22}
\bigskip\medskip
Reduction of Weak Interaction Rates in Neutron Stars by Nucleon Spin
Fluctuations: Degenerate Case}
\author{Georg Raffelt and Thomas Strobel}
\address{Max-Planck-Institut f\"ur Physik,
F\"ohringer Ring 6, 80805 M\"unchen, Germany}
\date{Received: 08 April 1996}
\maketitle
\begin{abstract}
Nucleon spin fluctuations in a dense medium reduce the ``naive''
values of weak interaction rates (neutrino opacities, neutrino
emissivities). We extend previous studies of this effect to the
degenerate case which is appropriate for neutron stars a few ten
seconds after formation.  If neutron-neutron interactions by a
one-pion exchange potential are the dominant cause of neutron spin
fluctuations, a perturbative calculation of weak interaction rates is
justified for $T\alt 3m/(4\pi\alpha_\pi^2)\approx 1\,\rm MeV$, where
$m$ is the neutron mass and $\alpha_\pi\approx15$ the pion
fine-structure constant.  At higher temperatures, the application of
Landau's theory of Fermi liquids is no longer justified, i.e.\ the
neutrons cannot be viewed as simple quasiparticles in any obvious
sense.
\end{abstract}
\vskip2.2pc]


\section{Introduction}

In a dense nuclear medium the effective neutrino interaction rates are
modified by the presence of nucleon-nucleon interactions. While the
importance of spatial spin-spin correlations has been recognized for a
long time, it had been overlooked that the interaction-induced
temporal fluctuations of the spin of a single nucleon can be a more
important effect.  It reduces the naive neutrino opacities and
neutrino emissivities of nuclear matter below their naive values
\cite{Raffelt,RS95,RSS,Sawyer95}.  These studies focussed on a
classical nucleon plasma, i.e.\ on nonrelativistic and nondegenerate
conditions which are thought to obtain in the core of a supernove for
the first few seconds after collapse. It was found that the 
spin-fluctuation rate in this environment is so large that it is not
possible to calculate weak interaction rates by a perturbative
expansion in terms of the nucleon-nucleon interaction potential.

We presently study the same effect for a degenerate medium in order to
derive a perturbative expression for the cross-section reduction by
nucleon spin fluctuations, and in order to understand the physical
conditions of temperature and density where a ``naive'' calculation of
weak interaction rates may be justified.  Many attempts have been made to
calculate neutrino opacities and emissivities for the physical
conditions pertaining to neutron stars because of the obvious
importance of these quantities for a theoretical understanding of
neutron-star cooling \cite{Reactions}.  While many of these works are
dedicated to calculating the impact of spatial correlation effects on
neutrino interaction rates, none of them appears to have addressed the
important issue of nucleon spin autocorrelations.

One may take a somewhat different perspective on the same problem if
one notes that these calculations are based on Landau's theory of
interacting Fermi liquids where a ``nucleon'' is a
quasiparticle excitation of the medium \cite{Landau}.  This picture is
justified only if the quasiparticles near the Fermi surface do not
interact too strongly, i.e.\ $\tau^{-1}\ll T$, where $\tau$ is a
typical time between collisions and $T$ is the temperature of the
medium.  Landau's condition is based on the observation that at $T=0$
the Fermi-Dirac distribution is a step function which, at nonzero
temperature, is smeared out over an energy range of approximate width
$T$. Collisions, on the other hand, introduce an energy uncertainty of
order $\tau^{-1}$ which clearly should be much smaller than $T$ in
order for the Fermi-Dirac distribution to make any sense. When
Landau's condition is violated it is not possible 
to speak of quasiparticles which obey Fermi statistics.  Degeneracy
effects ensure that the time between collisions becomes large at low
temperatures, so there is no significant restriction in the $T\to 0$
limit. For hot neutron-star matter, however, it is not {\it a priori}
obvious that Landau's condition is satisfied.  We were unable to
locate any discussion of this problem in the entire literature
pertaining to weak interaction rates in neutron stars
\cite{Reactions}.  Therefore, it is not frivolous to raise the
question of how cold the medium in a neutron star has to become before
a Fermi-liquid treatment becomes possible.

As previously argued \cite{Raffelt,RS95,RSS}, the cross-section
reduction by nucleon spin fluctuations becomes large when a typical
nucleon spin-fluctuation rate is of order the ambient temperature $T$
or larger. Since nucleons interact by a spin-dependent force, the
spin-fluctuation rate is roughly identical with the nucleon collision
rate.  Therefore, the condition that the spin-fluctuation rate be much
less than $T$ ensures both that the weak interaction rates are not
much affected by nucleon spin fluctuations and that Landau's condition
is satisfied.

The main problem in the degenerate case is to identify the quantity
which is to be interpreted as the relevant effective spin-fluctuation
rate. Because only the spins of nucleons near the Fermi surface have a
chance of evolving in a nontrivial way, and because Landau's condition
pertains to the quasiparticles near the Fermi surface, it is clear
that we need to define an appropriate effective spin-fluctuation rate
for the quasiparticles near the Fermi surface.

The impact of nucleon-nucleon collisions on weak interaction rates is
best understood in the language of linear-response theory where the
medium is described by the dynamical density and (iso)spin-density
structure functions. This method allows for a straightforward
calculation of the reduction of weak interaction rates in the
perturbative limit where the Landau condition is fulfilled, and thus
allows for a delineation of the physical parameters where this
treatment is justified.  

We will limit ourselves to the simple situation of a nonrelativistic,
single-species medium, i.e.\ we will study nonrelativistic degenerate
neutron matter. This excludes the important Urca processes from
consideration which are more difficult to address because they involve
two degenerate Fermi seas (protons and neutrons) with vastly different
Fermi momenta.  We believe that for the present exploratory purposes a
simple toy model is best suited to illuminate the issues at
hand. Therefore, we shall limit our attention to the neutral-current
scattering process $\nu+n\to n+\nu$ in nonrelativistic degenerate
neutron matter in the presence of interactions which cause the neutron
spins to fluctuate.

In Sec.~II we introduce the relevant spin-density structure function
and derive a simple sum rule which is used in Sec.~III to calculate
the $\nu n$ scattering cross-section reduction by neutron spin
fluctuations. In Sec.~IV we summarize and discuss our result.


\section{Dynamical Structure Functions} 

\subsection{Definition}

In nonrelativistic neutron matter all weak
interaction rates are determined by the dynamical density and 
spin-density structure functions. In an isotropic medium they are 
given by \cite{RS95}
\begin{eqnarray}\label{E002}
S_\rho(\omega,{\bf k})&=&\frac{1}{n_BV}\int_{-\infty}^{+\infty} dt\,
e^{i\omega t}\langle\rho(t,{\bf k})\rho(0,-{\bf k})\rangle,\cr
\noalign{\vskip3pt plus1pt}
S_\sigma(\omega,{\bf k})&=&\frac{4}{3n_BV}\int_{-\infty}^{+\infty} 
dt\,
e^{i\omega t}\langle\hbox{\boldmath$\sigma$}(t,{\bf k})\cdot
\hbox{\boldmath$\sigma$}(0,-{\bf k})\rangle,
\end{eqnarray}
where $n_B$ is the baryon (here neutron) density, $V$ the volume of
the system, $\bf k$ the
momentum transfer, and $\omega$ the energy transfer from the weak
probe to the neutron medium. Further, $\rho(t,{\bf k})$ is the spatial
Fourier transform at time $t$ of the neutron density operator
$\rho(x)=\psi^\dagger(x)\psi(x)$ where $\psi(x)$ is the neutron field
operator, a Pauli two-spinor. Similarly,
$\hbox{\boldmath$\sigma$}(t,{\bf k})$ is the Fourier transform of the
spin-density operator $\hbox{\boldmath$\sigma$}(x)=\frac{1}{2}
\psi^\dagger(x)\hbox{\boldmath$\tau$}\psi(x)$ with
$\hbox{\boldmath$\tau$}$ a vector of Pauli matrices. 
The expectation value $\langle\ldots\rangle$ is taken over a thermal
ensemble so that detailed balance 
$S_{\rho,\sigma}(\omega,{\bf k})=S_{\rho,\sigma}(-\omega,-{\bf k})
e^{\omega/T}$ is satisfied. Note that a positive $\omega$ is energy
given to the medium by the weak probe. 

The energies of the neutrinos which interact with the medium are much
smaller than the neutron mass or momenta so that the
long-wavelength limit ${\bf k}\to 0$ is an adequate first
approximation. In practice, its validity is questionable if
neutron-neutron correlations or collective modes are important which
for the moment we shall assume is not the case. Then the neutrino
differential scattering cross section is given by 
\begin{equation}\label{E003}
\frac{d\sigma}{d\epsilon_2}=\frac{G_{\rm F}^2\,\epsilon_2^2}{4\pi}\,
\left(C_V^2\,\frac{S_\rho(\epsilon_1-\epsilon_2)}{2\pi}+
3C_A^2\,\frac{S_\sigma(\epsilon_1-\epsilon_2)}{2\pi}\right),
\end{equation}
where $\epsilon_{1,2}$ is the energy of the incoming and outgoing
neutrino, respectively, and $S_{\rho,\sigma}(\omega)$ stands for
$\lim_{{\bf k}\to0}S_{\rho,\sigma}(\omega,{\bf k})$.  Further, $G_{\rm
F}$ is the Fermi constant, and $C_V=-1$ and $C_A\approx-1.15$ are the
neutral-current weak coupling constants for the neutron \cite{RS95}.
In bulk nuclear matter, $C_A$ may be suppressed somewhat.

In a noninteracting medium, the density and spin-density operators
remain constant so that the dynamical structure functions are
proportional to $\delta(\omega)$. In the nondegenerate case, they are
$S_{\rho,\sigma}(\omega)=2\pi\delta(\omega)$. To include
neutron-neutron anticorrelations induced by the Pauli exclusion
principle one evaluates the expectation values in Eq.~(\ref{E002}) by
normal ordering of the neutron field operators, taking proper account
of the anticommutation relations. Then one arrives at the intuitive
result
\begin{equation}\label{E005}
S_{\rho,\sigma}(\omega)=2\pi\,\delta(\omega)\,
\frac{1}{n_B}\int\frac{2d^3{\bf p}}{(2\pi)^3}\,
f_{\bf p}(1-f_{\bf p}),
\end{equation}
where $f_{\bf p}$ is the occupation number of the neutron field mode
${\bf p}$. 
In the nondegenerate limit one may neglect the Pauli blocking factor 
$(1-f_{\bf p})$ so that one arrives at the previous result if one 
notes that $n_B=\int f_{\bf p} 2d^3{\bf p}/(2\pi)^3$. Here, the factor
2 counts the two neutron spin degrees of freedom.

Even after ``turning on'' interactions between the neutrons, or
between the neutrons and some external potential, the density operator
remains constant. The vector current quantity that does fluctuate in
the presence of interactions is the neutron velocity which in the
nonrelativistic limit is small. Therefore, $S_\rho(\omega)$ remains
proportional to $\delta(\omega)$.

However, if the interaction involves a spin-dependent force as
expected for neutron-neutron interactions, the spin-density structure
function will be broadened because the spin of a given neutron near
the Fermi surface will ``forget'' its initial orientation roughly
after the collision time $\tau$. The width of $S_\sigma(\omega)$
roughly represents $\tau^{-1}$ so that the Landau condition
corresponds to the requirement that the width of $S_\sigma(\omega)$
must be much less than $T$.  If this is satisfied, the neutrino
scattering rates and thus the neutrino opacities are well approximated
by the noninteracting result for $S_\sigma(\omega)$.  Of course, it
may be modified by neutron-neutron correlations or collective modes,
effects that were the main focus of many of the previous
papers~\cite{Reactions}.


\subsection{Normalization}

An important general property of the dynamical structure functions is
their normalization. If one integrates both sides of Eq.~(\ref{E002})
over $d\omega$, the term $e^{i\omega t}$ yields $\delta(t)$ so that
the time integral can be trivially done. Then the normalization for
the spin-density case is
\begin{equation}\label{E009}
\int_{-\infty}^{+\infty}\frac{d\omega}{2\pi}\,
S_\sigma(\omega)=\frac{4}{3n_BV}\,
\langle\hbox{\boldmath$\sigma$}(0,0)\cdot
\hbox{\boldmath$\sigma$}(0,0)\rangle.
\end{equation}
If one ignores spin-spin correlations, the r.h.s.\ is independent of
the neutron spins' evolution. For the sake of argument one may imagine
that this evolution is caused by the interaction with some external
potential rather than by neutron-neutron collisions so that there is
no reason to expect spin-spin correlations.

In this case one may evaluate the r.h.s.\ of Eq.~(\ref{E009}) as above
and finds
\begin{equation}\label{E010}
\int_{-\infty}^{+\infty}\frac{d\omega}{2\pi}\,
S_{\rho,\sigma}(\omega)=
\frac{1}{n_B}\int\frac{2d^3{\bf p}}{(2\pi)^3}\,
f_{\bf p}(1-f_{\bf p}).
\end{equation}
The occupation numbers are given by a Fermi-Dirac 
distribution so that the r.h.s.\ is
\begin{equation}\label{E011}
\frac{1}{n_B}\int\frac{2d^3{\bf p}}{(2\pi)^3}\,
\frac{1}{e^{(E-\mu)/T}+1}\left(1-\frac{1}{e^{(E-\mu)/T}+1}\right),
\end{equation}
where $E={\bf p}^2/2m$ is the kinetic energy, 
$m$ the neutron quasiparticle effective
mass, and $\mu$ the nonrelativistic neutron chemical
potential. Then Eq.~(\ref{E011}) is 
\begin{equation}\label{E012}
\frac{1}{n_B\pi^2}\int_{0}^\infty dp\, p^2\,
\frac{e^z}{(e^z+1)^2},
\end{equation}
where $z\equiv (E-\mu)/T$. For very degenerate conditions the
integrand is strongly peaked near $z=0$ (the Fermi surface) so that
after a transformation of the integration variable to $z$
one may replace $p$ with $p_{\rm F}$ 
and one may extend the lower limit of integration to
$-\infty$. The integral can then be evaluated analytically so that
altogether
\begin{equation}\label{E013}
\int_{-\infty}^{+\infty}\frac{d\omega}{2\pi}\,S_{\rho,\sigma}(\omega)=
\frac{3}{2\eta}.
\end{equation}
Here,
\begin{equation}\label{E014}
\eta=\frac{E_{\rm F}}{T}=\frac{p_{\rm F}^2}{2 m T}
\end{equation}
is the degeneracy parameter in the nonrelativistic and very 
degenerate limit with $E_{\rm F}=p_{\rm F}^2/2m$ the nonrelativistic
Fermi energy. 
   
Therefore, in a noninteracting degenerate 
medium the structure functions are
$S_{\rho,\sigma}(\omega)=(3/2\eta)\,2\pi\delta(\omega)$. The
total scattering cross section of a neutrino with energy $\epsilon_1$
is then $\sigma=(3/2\eta)\,
(C_V^2+3C_A^2)\,G_{\rm F}^2\,\epsilon_1^2/4\pi$.


\subsection[...]{Perturbative Representation of 
                 {\boldmath$S_\sigma(\omega)$}}

If neutrons interact by a spin-dependent force it causes a nontrivial
evolution of their spins and thus a nonzero width of
$S_\sigma(\omega)$. Except in the neighborhood of $\omega=0$ where
multiple-scattering effects become important, $S_\sigma(\omega)$ can
be calculated on the basis of a bremsstrahlung or medium-excitation
amplitude \cite{RSS,Sawyer95}. Because for small $\omega$ the result
generically varies as $\omega^{-2}$ it is useful to represent it in
the form
\begin{equation}\label{E015}
S^{\rm brems}_\sigma(\omega)=
\frac{\Gamma_\sigma}{\omega^2}\,s(\omega/T)\times
\cases{e^{\omega/T}& for $\omega<0$,\cr
                  1& for $\omega>0$.\cr}
\end{equation}
The explicit distinction between positive and negative energy
transfers represents the detailed-balance condition.  Further, $s(x)$
is an even function which is normalized such that $s(0)=1$. In the
classical limit of hard collisions one has $s(x)=1$ for all $x$ as
discussed in more detail in Refs.~\cite{RSS,Raffelt96}.  In general,
$s(x)$ embodies information about the detailed form of the interaction
potential and about quantum corrections to the classical result.  In
the nondegenerate case, $\Gamma_\sigma$ has the interpretation of an
average spin rate of change, or conversely, $\Gamma_\sigma^{-1}$ is
the approximate time for a given nucleon spin to relax, i.e.\ to
forget its initial orientation.

Explicit calculations of $\Gamma_\sigma$ and $s(x)$ exist for a
single-species nuclear medium where the nucleon interaction is
modelled by a one-pion exchange (OPE) potential~\cite{RS95}. For a  
degenerate medium the relevant expressions can be
extracted from Ref.~\cite{FrimanMaxwell}
\begin{equation}
\label{E018}
\Gamma_{\sigma,{\rm OPE}}=
4\pi\alpha_\pi^2 T^3/p_{\rm F}^2,
\end{equation}
where the neutron Fermi momentum is given by $n_B=p_{\rm F}^3/3\pi^2$,
$\alpha_\pi\equiv (f 2m/m_\pi)^2/4\pi\approx15$ with
$f\approx1.0$ is the pion fine-structure constant, $m$ is
the neutron mass, and the pion mass has been neglected in the OPE
potential. One also finds from Ref.~\cite{FrimanMaxwell}
\begin{equation}
\label{E019} 
s_{\rm OPE}(x)=\frac{(x^2+4\pi^2)\,|x|}{4\pi^2(1-e^{-|x|})}\,,
\end{equation}
which is 1 at $x=0$ while for $|x|\gg 1$ it is
$|x|^3/4\pi^2$. 

Sigl~\cite{Sigl} has derived an f-sum rule which implies that the
integral $\int S_\sigma(\omega)\,\omega\,d\omega$ must exist and thus
that $s(x)$ must be a decreasing function for large $x$. This
conclusion also pertains to the degenerate case: if the energy
transfer $\omega$ to the medium far exceeds the
Fermi energy $E_{\rm F}$, a nucleon is lifted far above the Fermi
surface so that degeneracy effects cannot cause a modification of the
nondegenerate result.  Thus, the degenerate and nondegenerate $s(x)$
must be identical for $|x|\gg E_{\rm F}/T$ apart from a
multiplicative factor which arises because of our normalization
$s(0)=1$. 

Explicit calculations of $s(x)$ for various assumptions
concerning the neutron interaction potential and for various degrees
of neutron degeneracy are left for a future study \cite{Strobel}. 


\subsection[...]{Physical Estimate of {\boldmath$\Gamma_\sigma$}}

It will turn out that the $\nu n$ scattering cross-section reduction
is primarily sensitive to the neutron spin-fluctuation rate
$\Gamma_\sigma$.  Therefore, it is useful to understand on physical
grounds its overall magnitude and its scaling with temperature and
density. To this end we assume that neutrons scatter with a
velocity-independent cross section $\sigma_n$ which is caused by a
spin-dependent force such that the neutron spin is flipped in a
typical collision. If the interaction is approximated by an OPE
potential, on dimensional grounds the cross section is estimated to be
$\sigma_n\approx\alpha_\pi^2/m^2$.  We will
assume that the scattering is either due to a random collection of
external scattering centers with a density $n_c$, or due to collisions
with the other neutrons with a density $n_B$.

If the neutrons are nondegenerate they move with a typical thermal
velocity $v\approx (3T/m)^{1/2}$. By assumption the spin-fluctuation
rate is roughly equivalent to the collision rate so that
$\Gamma_\sigma\approx n_c \langle \sigma_n v\rangle\approx n_c\sigma_n
(3T/m)^{1/2}$. With the above estimate for $\sigma_n$ and with the
other neutrons being the scattering centers ($n_c=n_B$) one finds that
 $\Gamma_\sigma$ scales as $\alpha_\pi^2 T^{1/2}
m^{-5/2}$. This agrees with an explicit calculation which yields
$4\sqrt\pi$ for the numerical factor \cite{RS95}.

Next, we consider degenerate neutrons for which a typical velocity is
$p_{\rm F}/m$. If they interact with external scattering centers, the
collision rate for neutrons near the Fermi surface is about 
$n_c\sigma_n p_{\rm F}/m$. However, only the scattering of neutrons 
with an energy $E$ within about a distance $T$ from the Fermi surface
is not blocked by degeneracy effects. This is an approximate fraction
$T/E_{\rm F}=1/\eta$ of all neutrons. Therefore, the spin-fluctuation
rate averaged over all neutrons is $\Gamma_\sigma\approx n_c \sigma_n
(p_{\rm F}/m) (T/E_{\rm F})$.

Finally, if the scattering is among degenerate neutrons we have
$n_c=n_B$ and a typical relative velocity $p_{\rm F}/m$. The average
collision rate among neutrons is reduced by several factors of
$\eta^{-1}=T/E_{\rm F}$. Two such factors arise because each
initial-state neutron must have an energy within about $T$ of the
Fermi surface. One further factor arises because one final-state
neutron must also lie near the Fermi-surface; energy-momentum
conservation then ensures that the other one fulfills this condition
as well. Altogether we thus find $\Gamma_\sigma\approx \sigma_n n_B
(p_{\rm F}/m) (T/E_{\rm F})^3$. With $n_B=p_{\rm F}^3/3\pi^2$ and
$\sigma_n\approx\alpha_\pi^2/m^2$ we thus recover Eq.~(\ref{E018})
apart from the numerical coefficient. This $\Gamma_\sigma$ is the
spin-fluctuation rate averaged over all neutrons.  The
spin-fluctuation rate of those neutrons which lie near the Fermi
surface is larger by a factor $\eta$.


\section{Cross-Section Reduction}

\subsection{General Result}

We may now proceed to calculate the $\nu n$
scattering cross section in the presence of spin fluctuations of the
degenerate neutrons.  To this end we begin with the total
axial-current scattering cross section $\sigma_A$ of a neutrino with
energy $\epsilon_1$.  In the structure-function language it is the
$d\varepsilon_2$ integral of the axial part of Eq.~(\ref{E003}) or
equivalently
\begin{equation}
\sigma_A=\frac{3C_A^2G_{\rm F}^2}{4\pi}\,
\int_{-\infty}^{+\infty} \frac{d\omega}{2\pi}\,S_\sigma(\omega)\,
(\varepsilon_1-\omega)^2\,\Theta(\varepsilon_1-\omega).
\label{III.1}
\end{equation}
The problem with this expression is that it diverges if one uses the
perturbative expression $S^{\rm brems}_\sigma(\omega)$ instead of the
full but unknown $S_\sigma(\omega)$.  Following the treatment of the
nondegenerate case \cite{RSS} we note, however, that Eq.~(\ref{III.1})
can still be evaluated on the basis of $S^{\rm brems}_\sigma(\omega)$
without knowledge of the detailed low-$\omega$ behavior if one
includes the sum rule Eq.~(\ref{E013}).

To this end we note that for degenerate free neutrons 
the $\nu n$ scattering cross section
is $\sigma_{A,\rm free}= 
(3/2\eta)\,(3/4\pi)\,C_A^2G_{\rm F}^2\,\varepsilon_1^2$ 
as stressed after Eq.~(\ref{E014}). Therefore,
the interaction-induced modification $\delta\sigma_A\equiv
\sigma_{A}-\sigma_{A,\rm free}$ is given by
\begin{equation}
\frac{\delta\sigma_A}{\sigma_{A,\rm free}}
=-1+\int_{-\infty}^{+\infty} \frac{d\omega}{2\pi}\,
\frac{2\eta\,S_\sigma(\omega)}{3}\,
\frac{(\varepsilon_1-\omega)^2\,\Theta(\varepsilon_1-\omega)}
{\varepsilon_1^2}\,.
\label{III.2}
\end{equation}
Then we may proceed analogously to Ref.~\cite{RSS} and replace $-1$
with an integral over the structure function by virtue of the sum rule
Eq.~(\ref{E013}),
\begin{equation}
\frac{\delta\sigma_A}{\sigma_{A,\rm free}}=
\int_{-\infty}^{+\infty} \frac{d\omega}{2\pi}
\frac{2\eta\,S_\sigma(\omega)}{3}
\left[\frac{(\varepsilon_1-\omega)^2\Theta(\varepsilon_1-\omega)}
{\varepsilon_1^2}-1\right].
\label{III.3}
\end{equation}
For small $\omega$ the integrand varies effectively as
$\omega^2S_\sigma(\omega)$ because the term linear in $\omega$
switches sign at the origin. Therefore, to lowest order we may
substitute $S_\sigma(\omega)\to S^{\rm brems}_\sigma(\omega)$,
provided we interpret the remaining integral by its principal part.

This result becomes more transparent if we consider the reduction of
an average cross section rather than one for a fixed initial-state
neutrino energy. To this end we use a Maxwell-Boltzmann distribution
of neutrino energies at the same temperature $T$ which characterizes
the ambient neutron medium. The thermally averaged free cross section
is found to be $\langle\sigma_{A,\rm free}\rangle=
(3/2\eta)\,(9/\pi)\,C_A^2 G_{\rm F}^2 T^2$.  Because
Eq.~(\ref{III.3}) is fully analogous to the corresponding result of
Ref.~\cite{RSS} apart from an overall factor $2\eta/3$ we may conclude
without further calculations that
\begin{equation}
\frac{\delta\langle\sigma_A\rangle}
{\langle\sigma_{A,\rm free}\rangle}=
-\frac{2\eta}{3}\int_0^\infty \frac{dx}{2\pi}\,
\widetilde S_\sigma(x)\,G(x),
\label{III.4}
\end{equation}
where $x=\omega/T$,
\begin{equation}
G(x)=1-(1+x+{\textstyle\frac{1}{6}}\,x^2)\,e^{-x}=
\textstyle{\frac{1}{3}} x^2 +{\cal O}(x^3),
\label{III.5}
\end{equation}
and $\widetilde S_\sigma(x)\equiv T S_\sigma(x T)$.

As in Ref.~\cite{RSS} the $x^2$ behavior of $G(x)$ at small $x$ allows
us to replace $\widetilde S_\sigma(x)$ to lowest order with the
perturbative $\widetilde S^{\rm brems}_\sigma(x)$. Therefore, with the
representation Eq.~(\ref{E015}) and with
$\gamma_\sigma\equiv\Gamma_\sigma/T$ we find for the cross-section
reduction
\begin{equation}
\frac{\delta\langle\sigma_A\rangle}
{\langle\sigma_{A,\rm free}\rangle}=
-\frac{2\eta}{3}\,\frac{\gamma_\sigma}{2\pi}
\int_0^\infty dx\,x^{-2}G(x)\,s(x).
\label{III.6}
\end{equation}
The integral expression is $5/6$ for the classical approximation
$s(x)=1$. In general, the integral will be a numerical expression of
order unity. Its precise value for a variety of assumptions concerning
the cause for the neutron spin fluctuations will be studied elsewhere
\cite{Strobel}. 

Equation~(\ref{III.6}) shows that the expansion parameter which
defines the perturbative regime is $2\eta\gamma_\sigma/3$, as opposed
to the nondegenerate case where it was found to be $\gamma_\sigma$. In
both cases $\gamma_\sigma$ is defined to be the spin-fluctuation rate
averaged over all neutrons of the medium. However, in the degenerate
case only the neutrons near the Fermi surface participate in
collisions; it is {\it their\/} spin-fluctuation rate which reduces
the $\nu n$ scattering cross section.  The quantity
$2\eta\gamma_\sigma/3$ corrects for this effect. It is to be
interpreted as an effective spin-fluctuation rate for the neutrons
near the Fermi surface, in agreement with our estimates of Sec.~II.D.

We conclude that a ``naive'' perturbative calculation of neutrino
interaction rates in a degenerate neutron medium is possible if
$\eta\Gamma_\sigma\ll T$ while significant correction arise if
$\eta\Gamma_\sigma\agt T$. This latter case corresponds to a situation
where the collision rate of neutrons near the Fermi surface is not
small relative to $T$, in violation of Landau's condition for the
applicability of a Fermi-liquid treatment.  

In the nondegenerate case it was reasonable to extra\-polate the
behavior of the cross section $\langle \sigma_A\rangle$ into the
nonperturbative regime by virtue of an explicit ansatz for the
low-$\omega$ behavior of $S_\sigma(\omega)$ which incorporated the
equivalent of the sum rule Eq.~(\ref{E013}). In the present case such
an extrapolation is far more problematic because the derivation of the
sum rule itself was based on the assumption that neutrons can be
treated as quasiparticles which follow a thermal Fermi-Dirac
distribution. In the nonperturbative regime this assumption is not
justified so that in the present case the sum rule has a weaker
standing than it did in the nondegenerate case where we did not need
to invoke the anticommutation relations for the nucleon fields. 


\subsection{Numerical Result for OPE Potential}

If neutron-neutron collisions are the primary cause for neutron spin
fluctuations, and if one models the interaction by an OPE potential,
we may use Eq.~(\ref{E018}) to estimate $\Gamma_\sigma$. In this case
we find
\begin{equation}
\label{III.7}
\frac{2\eta}{3}\,\frac{\Gamma_{\sigma,\rm OPE}}{T}
=\frac{4\pi}{3}\,\alpha_\pi^2\,\frac{T}{m}
=1.00\,\frac{T}{\rm MeV},
\end{equation}
where we have used the vacuum neutron mass for the numerical estimate.
This result does not depend on the density (or Fermi momentum) which
fortuitously cancels as explained by the physical arguments in
Sec.~II.D.  If the neutron spin fluctuations were caused by the
interaction with a distribution of external scattering centers,
$\Gamma_\sigma$ would depend on their density as well as on the
neutron Fermi momentum.

Of course, if neutron-neutron interactions are the primary cause for
neutron spin fluctuations one would also expect significant spin-spin
correlations which we have ignored. However, in order to study
spin-spin correlations in the framework of a Fermi liquid theory one
would need to assume that Landau's condition is fulfilled which is not
the case in any obvious sense when $\eta\Gamma_\sigma\agt T$.
Therefore, we believe that the usual calculations of neutrino
opacities in hot degenerate neutron-star matter are applicable only
for $T\alt 1\,\rm MeV$.


\section{Summary}

We have derived an expression for the $\nu n$ scattering cross-section
reduction in degenerate neutron matter caused by neutron spin
fluctuations. We have used the linear-response theory approach of
Ref.~\cite{RSS}, but undoubtedly one would reach the same result by
the direct perturbative method of Ref.~\cite{Sawyer95}.

In a neutron star, these spin fluctuations will be caused by a
spin-dependent $nn$ interaction potential. Therefore, in general
spin-spin correlations will also be important which may cause further
modifications of the scattering cross section. Many of the previous
papers which deal with weak interaction rates in neutron stars
\cite{Reactions} were dedicated to an analysis of these latter
effects. We stress, however, that these calculations were based on the
assumption that Landau's condition is satisfied which is roughly
equivalent to the requirement that the autocorrelation function of a
single nucleon spin near the Fermi surface is narrow on a scale set by
the ambient temperature $T$.

If neutron-neutron interactions are modelled by a one-pion exchange
potential we estimate that the usual perturbative calculations are
justified for $T\alt 1\,\rm MeV$, a temperature which is reached very
quickly in a neutron star after formation. Therefore, the long-term
cooling history remains unaffected. Of course, a calculation of the
long-term cooling history does not require knowledge of the neutrino
opacity anyway as at late times neutrinos are no longer trapped.
Roughly speaking, then, the neutrino opacities matter only for a short
period after formation of a neutron star.  However, precisely for this
period the $\nu n$ scattering rate cannot be calculated by
straightforward perturbative techniques on the basis of first
principles.


\section*{Acknowledgments}
We acknowledge partial support by the European Union contract
CHRX-CT93-0120 and by the Deutsche Forschungsgemeinschaft grant SFB
375.


\end{document}